\documentclass{ws-mplb}

\begin{document}

\markboth{A. Mourachkine}{Mechanism of high-$T_c$ superconductivity based 
mainly on tunneling measurements in cuprates}

%
\catchline{}{}{}{}{}
%

\title{MECHANISM OF HIGH-$T_c$ SUPERCONDUCTIVITY BASED MAINLY ON 
TUNNELING MEASUREMENTS IN CUPRATES\footnote{This brief review is based on 
the talk given at the University of Cambridge in May 2004, and the talk, in its 
turn, is based on a book (Ref. 6).}}

\author{\footnotesize A. MOURACHKINE}

\address{Cavendish Laboratory, University of Cambridge, Madingley Road,\\ 
Cambridge, CB3 0HE, UK\\  
andrei\_mourachkine@yahoo.co.uk}


\maketitle

\begin{history}
\received{June 2005}
\end{history}

\begin{abstract}
The main purpose of this {\em brief} review is to present a sketch of the 
mechanism of high-$T_c$ superconductivity based mainly on tunneling 
measurements in cuprates. In the review, we shall mostly 
discuss tunneling {\em spectroscopy} in Bi$_2$Sr$_2$CaCu$_2$O$_8$. 
Analysis of the data shows that the Cooper pairs in cuprates are topological 
excitations (e.g. quasi-one dimensional bisolitons, discrete breathers etc.), 
and the phase coherence among the Cooper pairs appears due to spin 
fluctuations. 
\end{abstract} 


\vspace*{2.5mm}
{\footnotesize PACS Number(s): 74.50.+r, 74.72.-h, 74.62.Dh, 72. 20.-z, 
72.25.Dw}

\section{Introduction} 

Superconductivity (SC) discovered in 1911 by Kamerlingh Onnes\cite{1} and 
his assistant Holst has remained a major scientific mystery for a large 
part of the last century. It was completely understood only in 1957 when 
Bardeen, Cooper and Schrieffer (BCS) formulated the microscopic theory of 
SC in metals.\cite{2} The central concept of the BCS theory is the weak 
electron-phonon interaction which leads to the appearance of an attractive 
potential between two electrons. As a consequence, some electrons form pairs, 
and thus, composite bosons. The long-range phase coherence among the pairs 
appears due to the overlap of their wavefunctions, leading to a 
peculiar correlated state of matter---a quantum state on a macroscopic 
scale, in which all the electron pairs move in a single coherent motion. 

In 1986, Bednorz and M\"uller found SC in La-Ba-Cu-O ceramics at 
30 K.\cite{3} In 1993, the maximum critical temperature of copper oxides 
(cuprates) reached 135 K. The crystal structure of cuprates is layered and 
highly anisotropic. The parent compounds of SC cuprates are 
antiferromagnetic (AF) Mott insulators. A Mott insulator has a charge gap of 
$\sim$ 2 eV, whereas the spin wave spectrum extends to zero energy. When 
cuprates are slightly doped by holes or electrons, on cooling they become SC. 
The doped carriers are accumulated in two-dimensional CuO$_2$ planes. 
Thus, for SC in cuprates, the carrier density in CuO$_2$ planes is the main 
factor. SC occurs at low temperatures when the doping level is 
approximately one doped carrier per three Cu$^{2+}$ ions ($\sim$ 16\%).

Soon after the discovery of high-$T_c$ superconductivity (HTSC), it became 
clear that the concept of the Fermi liquid is not applicable to cuprates: the 
normal state properties of cuprates are markedly different from those of 
conventional metals. A normal-state partial gap, the pseudogap (PG), appearing 
in electronic excitation 
spectra of cuprates above $T_c$, is one of the main features of HTSCs.\cite{4}  
There is an interesting contrast between the development of the physics of 
cuprates and that of the physics of conventional SCs. Just before the creation 
of the BCS theory, the normal-state properties of conventional metals were 
very well understood; however SC was not. The situation with the cuprates was 
just the opposite: at the time HTSC was discovered, there already existed a 
good understanding of the phenomenon of SC, but the normal-state properties 
of cuprates were practically unknown. 

In addition to their peculiar normal-state properties, several experiments 
show that some SC properties of cuprates deviate from predictions of the BCS 
theory. For example, the isotope effect is almost absent in {\em optimally} 
doped cuprates.\cite{5,R,R2} Thus, on a microscopic scale, there is a clear 
difference between conventional SCs and HTSCs, namely that they have 
a different origin and that different criteria are required for HTSCs than 
for classical SCs. In the absence of a commonly accepted theory of HTSC, 
it is still time to discuss experimental data obtained in cuprates. 

The main purpose of this brief review is to present a more or less coherent 
picture of the mechanism of HTSC based mainly on tunneling measurements in 
cuprates. In the review, we shall focus our attention on hole-doped cuprates. 
We shall mainly discuss tunneling {\em spectroscopy} in 
Bi$_2$Sr$_2$CaCu$_2$O$_8$ 
(Bi2212) performed by the present author. We shall also use some inelastic 
neutron scattering (INS) and muon spin resonance ($\mu$SR) data. The main 
ideas of the review appeared in 1998.\cite{8} 
Since the review is based on experimental data, the reader will not find many 
formulas in the text, not a single Hamiltonian. {\it Due to the restriction on the 
number of pages in a review, only limited number of experimental data can be 
presented here.} More tunneling data and a more detailed description of the 
mechanism of HTSC based on the data can be found elsewhere.\cite{R,R2} 

Analysis of the data shows that (i) there are two SC gaps in cuprates: a 
pairing and a phase coherence gap. (ii) Tunneling spectra below $T_c$ are a 
combination of coherent QP peaks and incoherent part from a normal-state 
PG. (iii) The ``tunneling'' normal-state PG is most likely a charge gap. (iv) 
However, SC cuprates have the second normal-state PG which is magnetic 
and, most likely, spatially separated from the charge PG. (v) Just above 
$T_c$, incoherent Cooper pairs contribute to the ``tunneling'' PG.    
(vi) There is a clear correlation between the magnitude of pairing gap and the 
magnitude of ``tunneling'' PG in Bi2212  ($\Delta_p \simeq \Delta_{pg}/3$). 
(vii) Tunneling characteristics corresponding to the SC condensate are in good 
agreement with theoretical predictions for topological excitations (solitons). 
(viii) In addition, analysis of INS and $\mu$SR data shows 
that the long-range phase coherence occurring at $T_c$ is established due 
to spin fluctuations into CuO$_2$ planes.

\section{Starting Point: Phase Diagram of Cuprates} 

Let us start the review with the ``basics," i.e. with what is widely accepted. 
Figure 1 shows a sketch of phase diagram of hole-doped cuprates. At low 
doping level, cuprates are AF. The SC state occurs at 
higher hole concentrations in CuO$_2$ planes. If in La$_{2-x}$Sr$_x$CuO$_4$ 
(LSCO), there is a distance between the AF and SC phases, in 
YBa$_2$Cu$_3$O$_7$ (YBCO), the SC state practically evolves from the AF 
phase. 
\begin{figure}[t]
\centerline{\psfig{file=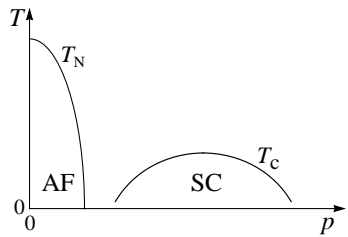,width=4.5cm}}
\vspace*{8pt}
\caption{Phase diagram of hole-doped cuprates: AF and SC phases. $T_N$ is 
the N\'eel temperature, and $p$ is the hole concentration into the CuO$_2$ 
planes.}
\end{figure}

In conventional SCs, the critical temperature rises monotonically with the rise 
of charge-carrier concentration, $T_c(p) \propto p$, where $p$ is  the carrier 
concentration. In cuprates, the $T_c(p)$ dependence is  nonmonotonic. In 
hole-doped cuprates, the $T_c(p)$ dependence has a  bell-like shape and, in most 
of them, can be approximated by the empirical expression  
$T_c(p) \simeq T_{c,max}[1 - 82.6(p - 0.16)^2]$, 
where $T_{c,max}$ is the maximum critical temperature for a given 
compound.\cite{9} 
SC occurs within the limits 0.05 $\leqslant p \leqslant  0.27$ which vary slightly 
in different cuprates. Different doping regions of the SC phase are mainly known 
as {\em underdoped} ($0.05 \leqslant p \leqslant 0.14$), {\em  optimally doped} 
($0.14 <  p < 0.18$) and {\em overdoped} ($0.18 \leqslant p \leqslant 0.27$). 
The insulating phase at $p <$ 0.05 is  usually called the {\em undoped} region. 
Above $p =$ 0.27, cuprates are 
practically metallic.  The state above  the SC dome is often called  a strange 
metal. Paradoxically, the $T_{c,max}$  in cuprates is located near $p$ = 0.16, 
whereas the SC condensation  energy has the maximum in the overdoped region 
near $p$ = 0.19.\cite{10}

The PG, a normal-state partial gap appearing in electronic excitation 
spectra of cuprates above $T_c$, is not shown in Fig. 1. The reason is very 
simple: there is no agreement on the {\em exact} doping dependence 
of the PG. In spite of the fact that there is a consensus on {\em general} 
doping dependence of the PG in hole-doped cuprates---the magnitude 
of the PG decreases with increase of hole concentration---however, 
there is a clear discrepancy between the phase diagrams inferred from transport, 
nuclear magnetic resonance (NMR) and heat-capacity measurements, on one 
hand, and from  tunneling measurements, on the other  hand. Transport, NMR 
and heat-capacity measurements  show that, in the overdoped  region, the PG is 
absent above $T_c$. At the same time, in tunneling measurements, the PG is 
observed well above $T_c$.  Generally speaking, the PG in cuprates may arise 
from charge-density waves (CDW), local  AF correlations [or spin-density waves 
(SDW)], or from their combination. Incoherent Cooper pairs may also induce a PG. 
From the common sense, however, the PG is most likely a charge gap because, 
at low doping level, cuprates are AF Mott insulators and, therefore,  a spin gap 
is absent at low $p$. Alternatively, there are two PGs in cuprates. 
\begin{figure}[t]
\centerline{\psfig{file=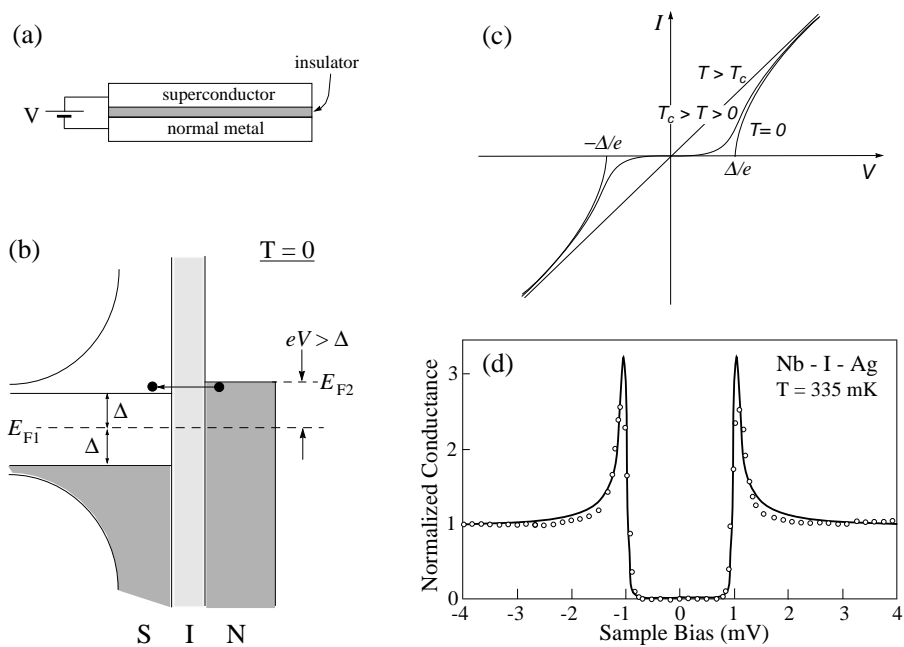,width=12cm}}
\vspace*{8pt}
\caption{(a) Sketch of an SIN tunneling junction, and (b) corresponding energy 
diagram at $T$ = 0 in the presence of an applied voltage: QPs can 
tunnel when $|V| \geq \Delta /e$. 
(c) Tunneling $I(V)$ characteristics for an SIN junction at 
different temperatures: $T$ = 0; 0 $< T < T_c$, and $T > T_c$ (the latter 
case corresponds to a NIN junction). At 0 $< T < T_c$, QP 
excitations exist at any applied voltage. 
(d) Tunneling conductance (circles) versus sample bias obtained in a 
Nb-I-Ag junction at 335 mK.\protect\cite{11} The solid line is 
a theoretical curve calculated from Eqs. (3) and (2).}
\end{figure}

\section{Tunneling Spectroscopy in Metals} 

The tunneling spectroscopy is a powerful tool, and has played a crucial role in 
the verification of the BCS theory.\cite{R} Tunneling spectroscopy is 
particularly sensitive to the density of state (DOS) near the Fermi level, $E_F$, 
and, thus, is capable of detecting any gap in the quasiparticle (QP) excitation
spectrum at $E_F$. In addition, it has a very high energy resolution: less than 
$k_B T$ for SC-insulator-SC (SIS) junctions, where 
$k_B$ is the Boltzmann constant. In comparison with photoemission 
measurements, tunneling spectroscopy has an additional advantage: 
to measure the DOS locally.  

\subsection{SIN tunneling}
  
Consider the flow of electrons across a thin insulating layer having the 
thickness of a few nanometers, which separates a normal metal from a 
conventional SC. Figure 2a shows a 
SC-insulator-normal metal (SIN) tunneling junction. At $T$ = 0, 
no tunneling current can appear if the absolute value of the applied voltage 
(bias) in the junction is less than $\Delta /e$. Tunneling will become 
possible when the applied bias reaches the value of $\pm \Delta /e$, as shown 
in Fig. 2b. Figure 2c shows schematically three current-voltage $I(V)$ 
characteristics for an SIN junction at $T$ = 0, 0 $< T < T_c$ and 
$T_c < T$. At $T$ = 0, the absence of a tunneling current at small voltages 
constitutes an experimental proof of the existence of a gap in the energy 
spectrum of a SC. At $0 < T < T_c$, there are always excited 
electrons due to thermal excitations. As shown in Fig. 2c, the $I(V)$ curves, 
obtained below $T_c$, approach at high bias the $I(V)$ characteristic 
measured above $T_c$ (i.e. Ohm's line). In conventional SCs, the gap completely 
vanishes at $T_c$. This, however, is not the case for cuprates which 
show the presence of a PG above $T_c$. 

\subsection{Density of states}  
 
In the framework of the BCS theory, the DOS of QP excitations in the 
SC state, $N_s(E)$, and the DOS in the normal state $N_n(E)$ relate to each 
other at $T$ = 0 as 
\begin{equation} 
N_s/N_n = \left\{ \begin{array}{ll} 
\frac{E}{\sqrt{E^{2} - \Delta ^{2}}} & \mbox{for $|E| \geq \Delta$} 
\nonumber \\  
0 & \mbox{if $|E| < \Delta$}. 
\end{array} \right. 
\end{equation} 
In most low-$T_c$ SCs, $N_n$ is normally constant over the energy range of 
interest. 

A $dI(V)/dV$ tunneling characteristic obtained in an SIN junction 
corresponds directly to the DOS of QP excitations 
in a SC. In a first approximation, assuming that the normal 
metal has a constant DOS near the Fermi level and the 
transmission of the barrier (insulator) is independent of energy, the 
tunneling conductance $dI(V)/dV$ is proportional to the DOS of 
a SC, broadened by the Fermi function 
$f(E,T) = [ \exp(E/k_{B}T) + 1]^{-1}$. Thus, at low temperature 
\begin{equation}
\frac{dI(V)}{dV} \propto \int\limits^{+ \infty}_{- \infty} N_s(E) 
\left[- \frac{\partial}{\partial (eV)} f(E + eV,T) \right]dE \cong N_s(eV), 
\end{equation}
where $e$ is the electron charge, and the origin of energy scale $E$ in 
the tunneling spectra corresponds to the Fermi level of the SC. 
Consequently, the differential conductance at negative (positive) voltage 
reflects the DOS below (above) $E_F$. 

In order to smooth the gap-related structures in the DOS 
$N_s \propto E/ \sqrt{E^{2} - \Delta ^{2}}$, a phenomenological smearing 
parameter $\Gamma$ was introduced, which accounts for a  lifetime broadening 
of QPs ($\Gamma = \hbar / \tau$, where 
$\tau$ is the lifetime of QP excitations).\cite{12} The energy $E$ in 
the DOS function is replaced by $E - i \Gamma$: 
\begin{equation}
N_s(E, \Gamma ) \propto Re \left\{ \int \frac{E - i \Gamma}{\sqrt{(E
-  i \Gamma )^2 - ( \Delta ( \vec{k} ))^2}} d \vec{k} \right\}, 
\end{equation}
where we introduced $\Delta ( \vec{k} )$ which is the $\vec{k}$-dependent 
energy gap for the general case of an anisotropic gap. In the two-dimensional 
case, the integration is reduced to the in-plane angle, 
0 $\leq \theta < 2 \pi$. 

In SIN tunneling junctions of conventional SCs, there is 
good agreement between the theory and experiment. As an example, 
Figure 2d shows the correspondence between experimental data obtained 
in Nb and the theoretical curve. 

\subsection{SIS tunneling}

In an SIS junction, the tunneling conductance is proportional to the convolution 
of the DOS function of a SC with itself. In the case 
of symmetrical SIS contacts, the expression for the tunneling current 
through contacts at finite temperature is 
\begin{equation}
I(V) = K \int\limits^{\infty}_0 N(E, \Gamma )N(E - eV, \Gamma ) 
[f(E,T) - f(E - eV,T)] dE, 
\end{equation}
where $K$ is the constant (matrix) which contains tunneling probabilities. 
The expression for tunneling conductance at finite temperature 
can be presented as 
\begin{equation}
\frac{(dI/dV)_s}{(dI/dV)_n} = \frac{d}{d(eV)} \int\limits^{eV}_0 
N(E, \Gamma )N(E - eV, \Gamma ) [f(E,T) - f(E - eV,T)] dE, 
\end{equation}
where $(dI/dV)_n$ is the conductance in the normal state. 

If, in SIN tunneling junctions of s-wave SCs, there is good 
agreement between the theory and experiment, as shown in Fig. 2d, in 
SIS junctions the correspondence between the BCS DOS and 
experimental data is poor (see p. 31 in Ref. 6). 

\section{Bi2212 Samples and Tunneling Techniques} 

Overdoped Bi2212 single crystals were grown by the self-flux method in 
Al$_2$O$_3$ and ZrO$_2$ crucibles and then mechanically separated from 
the flux. The dimensions of these crystals are typically 
2--3$\times1 \times$0.1 mm$^3$. The chemical composition of the  
Bi-2:2:1:2 phase in these single crystals corresponds to the formula 
Bi$_2$Sr$_{1.9}$CaCu$_{1.8}$O$_{8+x}$ as measured by energy 
dispersive X-ray fluorescence (EDX). The crystallographic $a$, $b$, 
$c$ values are of 5.41, 5.50 and 30.81 \AA, respectively. 
The $T_c$ value was determined by the four-contact method yielding 
$T_c$ = 87--90 K with the transition width less than 1 K. 
The underdoped Bi2212 single crystals were obtained by annealing the 
overdoped crystals in vacuum. 

Bi2212 single crystals in which Cu is partially substituted for Ni or Zn
were also grown by the self-flux method. As measured by EDX the chemical 
composition of the Bi-2:2:1:2 phase in Ni-doped and Zn-doped Bi2212 single 
crystals corresponds to the formula 
Bi$_2$Sr$_{1.95}$Ca$_{0.95}$(CuNi)$_{2.05}$O$_{8+x}$ and 
Bi$_2$Sr$_{1.98}$Ca$_{0.83}$(CuNi)$_2$O$_{8+x}$, respectively. 
In these single crystals the Ni and Zn content with respect to 
Cu was approximately 1.5\% and 1\%, respectively. The $T_c$ value was 
determined by the four-contact method: $T_c$ = 75--77 K in Ni-doped 
Bi2212 and $T_c$ = 76--78 K in Zn-doped Bi2212. In both types of 
single crystals, the transition width is a few degrees.  

Most tunneling data presented in this review were obtained by the 
break-junction (BJ) technique. Figure 3 schematically depicts the 
BJ setup. A single crystal is glued to a flexible insulating 
substrate, as shown in Fig. 3, such that the $ab$ plane of the crystal is 
perpendicular to the substrate surface. Four electrical contacts (typically 
with a resistance of a few Ohms) are made by attaching gold wires to 
the crystal by silver paint. The diameter of golden wires is 25 $\mu$m. 
Two contacts situated diagonally are used for bias input, which 
is slightly modulated by a lock-in, and the other two for measuring 
$dI(V)/dV$ and $I(V)$ tunneling characteristics. At low temperature, the 
crystal is broken in an He ambient in the $ab$ plane by bending the flexible 
substrate.  
The bending force is applied by the differential screw shown schematically 
in Fig. 3. The differential screw has a precision of 10 $\mu$m per turn. 
Tunneling is achieved by changing the distance between broken parts of the 
crystal. The $dI(V)/dV$ and $I(V)$ tunneling characteristics are determined 
by a standard lock-in modulation technique. 
\begin{figure}[t]
\centerline{\psfig{file=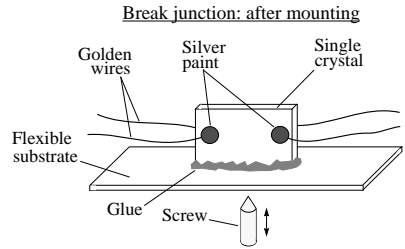,width=5.2cm}}
\vspace*{8pt}
\caption{Sketch of a BJ after mounting: a single crystal 
with four electrical contacts is glued to a flexible substrate. The crystal is 
broken by bending the substrate by a differential screw.}
\end{figure}

In one junction, a few tunneling spectra are usually obtained at low 
(constant) temperature by changing the distance between broken parts of 
a crystal, going back and forth etc., and each time the tunneling
occurs most likely in different places. 

In addition to SIS-junction measurements, tunneling tests have also been
carried out in Bi2212 single crystals by forming SIN junctions. 
In point-contact measurements, the differential screw shown in Fig. 3 
is used to push a normal tip against a fixed single crystal. Pt-Ir and Ag wires 
sharpened mechanically were used as normal tips. The point-contact 
measurements were performed either along or perpendicular to the $c$ 
crystal axis. 
\begin{figure}[t]
\centerline{\psfig{file=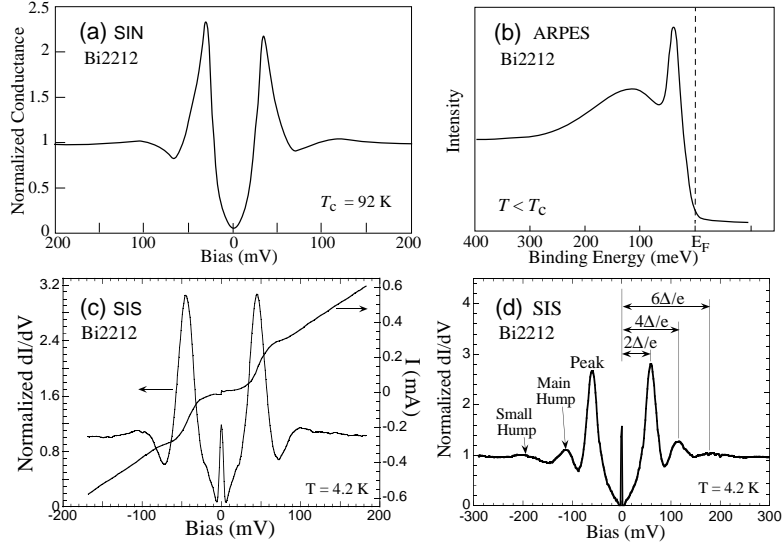,width=10.7cm}}
\vspace*{8pt}
\caption{(a) An averaged STM $dI/dV$ spectrum recorded in SIN junctions 
at 4.2 K in an optimally doped Bi2212 sample with $T_c$ = 92 
K.\protect\cite{13} 
(b) ARPES spectrum obtained in a slightly overdoped Bi2212 single 
crystal having $T_c$ = 91 K. $E_F$ is the Fermi level.\protect\cite{14}  
(c) Typical $dI/dV$ and $I(V)$ characteristics obtained in Bi2212 BJs. 
The data are recorded in a slightly overdoped Bi2212 single crystal. 
(d) A fine tunneling conductance obtained in a slightly overdoped 
Bi2212 single crystal with $T_c$ = 88 K in an SIS junction. The curve has 
QP peaks at $V = \pm2\Delta /e$, main humps at 
$\pm4\Delta /e$, and small humps at $\pm6\Delta /e$, where $\Delta$ 
is the pairing energy gap.\protect\cite{R}}
\end{figure} 

\section{Tunneling Spectroscopy in Bi2212}  

In this section, we discuss data obtained exclusively in {\em tunneling} regime, 
i.e. in junctions with a large normal resistance, $R_n \gg 0$. Let us 
start with SIN data. Figure 4a shows a conductance taken in an 
optimally doped Bi2212 single crystal. The conductance reflects typical 
features of Bi2212 spectra, namely, well-defined QP peaks, dips and 
humps outside the gap structure. The other feature of SIN conductances 
obtained in HTSC is that the height of the QP peaks is asymmetrical relatively 
zero bias: the QP peak at negative bias is always higher than that at positive 
bias (in SIS junctions, this is not the case). In SIN conductances, the humps are 
always situated at bias which is three times as large as that of the QP peaks, 
$|V_{hump}| \simeq 3\,|V_{peak}|$. Outside humps, the conductance is 
more or less constant.  

The magnitude of a SC gap can, in fact, be derived directly from the tunneling 
spectrum. However, in the absence of a generally accepted model for the gap 
function and the DOS in cuprates, such a quantitative analysis is not 
straightforward. Thus, in order to compare different spectra, we calculate 
the gap $\Delta$ magnitude in SIN junctions as a half spacing between the QP 
peaks at $V= \pm \Delta/e$. In Fig. 4a, the magnitude of energy gap is 
about 37 meV, while the humps are situate at about $\pm$110 mV. The gap 
inferred from conductances obtained in the {\em tunneling} regime will be 
referred to as the pairing gap, $\Delta_p$. 

Tunneling SIN data can directly be compared with angle-resolved photoemission 
spectroscopy (ARPES) data. Figure 4b shows an ARPES spectrum having the 
same peak-dip-hump structure as the SIN conductance in Fig. 4a. In the 
APRES spectrum, the QP peak is situated at about 40 meV, and the hump is 
at about 120 meV. 

Let us discuss now data obtained in SIS junctions. Figure 4c shows typical
$dI(V)/dV$ and $I(V)$ tunneling characteristics for SIS junctions of Bi2212. 
In addition to the peak-dip-hump structure, the conductance in Fig. 4c exhibits 
also a peak at zero bias due to the Josephson current. In SIS conductances, 
the distance between the peaks is $4\Delta_p/e$, the dips are more 
pronounced, and the humps appears at bias $V = \pm4\Delta_p/e$. As we 
shall see further, the humps in SIN conductance correspond to a normal-state 
gap, the tunneling PG. If so, then, in SIS spectra, the humps should be 
situated at bias of $V = \pm6\Delta_p/e$, and not at $\pm4\Delta_p/e$. 
As a matter of fact, there are humps in SIS spectra at bias of 
$\pm6\Delta_p/e$; however, they are very weak: they can be seen in 
Fig. 4d. The main humps in SIS conductances are in fact the superposition 
of ``QP peaks'' and ``true humps'' as illustrated in Fig. 5. This is 
the reason why the main humps in SIS spectra are situated at bias of 
$\pm4\Delta_p/e$. $I(V)$ characteristics of cuprates will be 
considered in detail further. 
\begin{figure}[t]
\centerline{\psfig{file=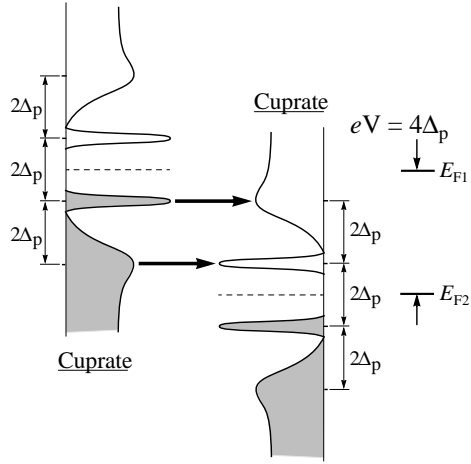,width=6.2cm}}
\vspace*{8pt}
\caption{Sketch of tunneling in an SIS junction of cuprates at $T$ = 0,  
explaining the origin of the main humps in SIS conductances which
appear at bias $V = \pm4\Delta_p /e$, shown in Figs. 4c, 4d and 6. 
The thick arrows indicate the tunneling process. $E_{F1}$ and $E_{F2}$ 
are the Fermi levels.\protect\cite{R}}
\end{figure}
 
As was mentioned in Introduction, for SC in cuprates, the carrier density 
in CuO$_2$ planes, $p$, is the main factor. The magnitude of pairing 
energy gap, $\Delta_p$, also depends on $p$. The dependence $\Delta_p(p)$ 
in Bi2212 is monotonic as shown in Fig. 6: the magnitude of $\Delta_p$
decreases as the doping level increases. Experimentally, the magnitude of 
$\Delta_p$ in Bi2212 linearly decreases as $p$ increases: 
$\Delta_p(p) \simeq 83.7\,[1 - \frac{p}{0.3}]$ (in meV).\cite{R} 
In Fig. 6a, one can see that there is a correlation between the magnitude 
of energy gap and the height and width of QP peaks: the smaller the gap 
magnitude is, the higher and narrower the QP peaks are. {\em However}, 
if one normalizes the conductances by height and by gap magnitude 
simultaneously, they all collapse in one curve (the peaks, {\em not the 
humps}) (see Fig. 12.47 in Ref. 6). 
\begin{figure}[t]
\centerline{\psfig{file=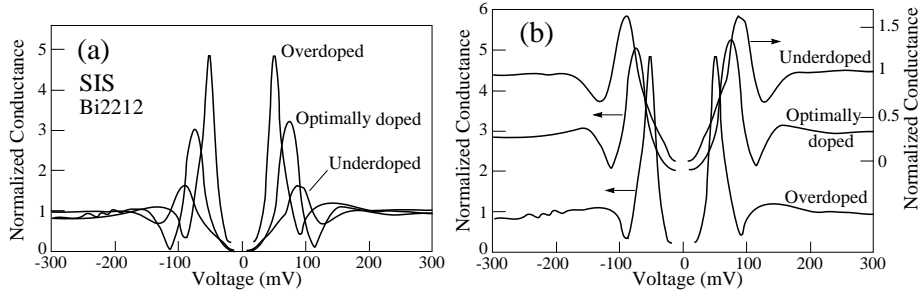,width=12.2cm}}
\vspace*{8pt}
\caption{(a) Tunneling conductances obtained in three Bi2212 single 
crystals by SIS junctions: 
in an underdoped Bi2212 with $T_c$ = 83 K, in an optimally 
doped Bi2212 having $T_c$ = 95 K, and in an overdoped Bi2212 
with $T_c$ = 82 K. In all curves, the Josephson 
current at zero bias has been removed. The curves are normalized at 200 mV. 
b) The same data as in plot (a) but the spectra are offset vertically for 
clarity.\protect\cite{15}} 
\end{figure}

\section{Two Superconducting Gaps: $\Delta_c$ and $\Delta_p$}  

Performing tunneling measurements not only in tunneling regime but also in 
junctions with small $R_n$, one will discover that there is
another energy scale in cuprates. Figure 7 shows two conductances obtained 
within the same Bi2212 single crystal by SIS junctions. 
One can see that these two spectra are different: they have different shapes 
and different magnitudes. The upper conductance has a subgap and a 
zero-bias conductance peak (ZBCP) due to the Josephson current, while the 
other does not. Instead, the lower curve is smooth between the QP peaks and 
has dips and humps outside the gap structure. 
These differences reflect the physics of the SC state in cuprates. The upper 
curve is measured in a junction with low $R_n$, while the lower curve in a junction 
with high $R_n$. These two conductances reflect the two energy gaps, 
$\Delta_c$ and $\Delta_p$,
where $\Delta_c$ is the phase coherence gap. Thus, $\Delta_c$ manifests 
itself in tunneling spectra {\em exclusively} in junctions with low $R_n$ 
($<$ 3 k$\Omega$). In contrast, $\Delta_p$ appears in high-$R_n$ junctions.  
Conductances corresponding to $\Delta_p$ always show the peak-dip-hump  
structure. In conductances corresponding to $\Delta_c$, the dips are either 
absent or very weak. In low-$R_n$ junctions, the ZBCP is usually large.  The 
two energy gaps are also present in\cite{17,R} YBCO and in electron-doped 
Nd$_{2-x}$Ce$_x$CuO$_4$ (NCCO).\cite{18,R} 
\begin{figure}[t]
\centerline{\psfig{file=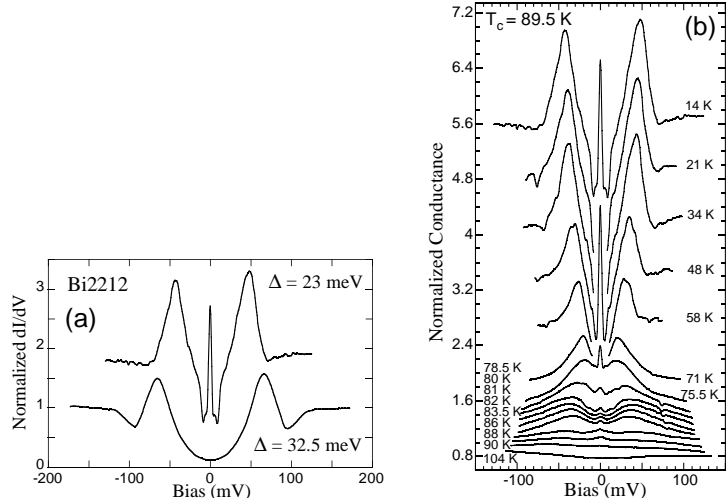,width=10cm}}
\vspace*{8pt}
\caption{(a) Two tunneling conductances obtained in a Bi2212 single crystal 
with $T_c$ = 89.5 K ($p \simeq$ 0.19) by SIS junctions.\protect\cite{R,8} 
The upper curve is offset vertically for clarity. The lower conductance is 
measured in a junction with $R_n \simeq$ 0.1 M$\Omega$. The upper 
conductance is recorded in a junction with $R_n \simeq$ 90 $\Omega$.
(b) Temperature dependence of the upper conductance in plot (a). 
The spectra are offset vertically for clarity. In the 
21 K, 34 K, 58 K and 71 K spectra, the Josephson current at zero bias has 
been removed for clarity.}
\end{figure} 
\begin{figure}[t]
\centerline{\psfig{file=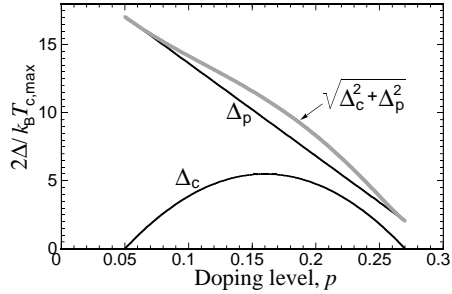,width=6cm}}
\vspace*{8pt}
\caption{Low-temperature phase diagram of SC cuprates: the pairing energy 
scale $\Delta_p$ is obtained in tunneling and ARPES measurements, while the 
phase-coherence energy scale $\Delta_c$ dominates in Andreev-reflection 
and penetration-depth  measurements.\protect\cite{R,R2,8,16}} 
\end{figure} 

The two gaps have different temperature dependences. 
Figure 7b depicts the temperature dependence of the upper conductance in 
Fig. 7a, which reflects the $\Delta_c(T)$ dependence. In Fig. 7b, one can see 
that, on heating, $\Delta_c$ vanishes at $T_c$. As a consequence, 
$\Delta_c(0) \propto T_c$ as in conventional SCs: 
$2\Delta_c(0) = \Lambda k_BT_c$, where $k_B$ is the Boltzmann constant,  
and the dependence $\Delta_c(p)$ has the shape of a dome, as shown in Fig. 8. 
This also means that the upper conductance in Fig. 7a, in a sense, corresponds 
to Andreev reflections. The coefficient $\Lambda$ slightly varies in different 
compounds: in Bi2212 $\Lambda \simeq 5.45$; in YBCO 
$\Lambda \simeq 5.1$, and in Tl$_2$Ba$_2$CuO$_6$ (Tl2201) 
$\Lambda \simeq 5.9$.\cite{R2} The dependence $\Delta_p(T)$ will be 
discussed 
further. The fact that the $\Delta_p$ energy scale in Fig. 8 lies above the SC 
dome indicates that $\Delta_p$ closes {\em above} $T_c$ at any doping level.  

The $\Delta_c$ energy scale manifests itself in {\em tunneling} 
measurements because always $\Delta_c < \Delta_p$. 
{\em In cuprates, it is easier to excite a Cooper pair than to break it.} The 
excited Cooper pairs can leave the SC condensate without being broken. In 
conventional SCs, this is impossible. This means that the two 
spectra in Fig. 7a reflect two types of tunneling: 
the upper conductance is caused by Cooper-pair tunneling, while the 
lower curve by single-electron tunneling. In order to extract a single 
electron from the SC condensate, the minimum energy 
2$\sqrt{\Delta _c^2 + \Delta _p^2}$ is required. For cuprates, this 
energy in not deferent much from $\Delta_p$, particularly in the 
underdoped region, as shown in Fig. 8. To break an uncondensed pair 
(e.g. on the surface), the energy $2\Delta_p$ is needed.  
It is important to note that ARPES studies\cite{14} do 
not detect the presence of $\Delta_c$; therefore, the $\Delta_c$ scale in 
Fig. 8 has exclusively the magnetic origin. There are no charges at 
$\Delta_c(p)$, they are ``located'' at the $\Delta_p(p)$ scale. 
\begin{figure}[t] 
\centerline{\psfig{file=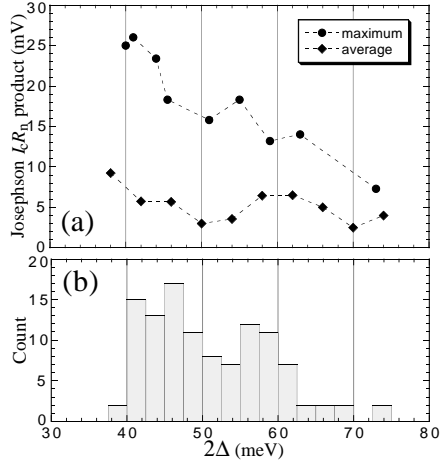,width=5.8cm}}
\vspace*{8pt}
\caption{(a) Maximum and average Josephson $I_cR_n$ products as 
functions of the gap magnitude in 110 Bi2212 break junctions. The doping 
level of Bi2212 single crystals is about 0.19 ($T_c \simeq$ 87--90 K). 
(b) The statistics of the gap magnitude for the  same 
110 junctions.\protect\cite{19,R}}
\end{figure}

Performing tunneling measurements in junctions with different $R_n$, one will 
find that, at any fixed $p$, there is a distribution of gap magnitude, in a 
first approximation, between $\Delta_c$ and 
$\sqrt{\Delta _c^2 + \Delta _p^2}$. The scale of the 
distribution depends also on the tunneling angle into and out of the CuO$_2$ 
planes. Figure 9b shows a distribution of the tunneling gap magnitude in 
Bi2212 at $p = 0.19$, which is in good agreement with the values of
2$\Delta _c \simeq$ 42 meV, 2$\Delta _p \simeq$ 61 meV, and 
2$\sqrt{\Delta _c^2 + \Delta _p^2} \simeq$ 74 meV, inferred from Fig. 8 
for Bi2212 at $p = 0.19$. We turn now to the correlation between the value 
of the Josephson product, $I_cR_n$, and the magnitude of the energy gap. 
Figure 9a depicts the maximum and the average $I_cR_n$ as a function of 
gap magnitude in Bi2212 at $p = 0.19$. Theoretically, in low-$T_c$ SCs, 
$I_cR_n$ is proportional to $\Delta$. In Fig. 9a, neither the average nor the 
maximum $I_cR_n$ are in agreement with the proportionality 
$I_cR_n \propto \Delta$. This is because, in cuprates, the smaller gap 
$\Delta_c$ is responsible for the phase coherence, and therefore, for the 
value of $I_cR_n$. In Fig. 9a, the average product is almost flat. 
This is due to the fact that, in cuprates, the value of $I_cR_n$ strongly 
depends on the surface quality in a junction. As a consequence, the 
average $I_cR_n$ does not reflect the real $I_cR_n$ values. However, by doing 
a statistical study, the real value of $I_cR_n$ can occasionally be observed or, 
at least, the values close to the actual ones. In Fig. 9a, the maximum $I_cR_n$ 
almost linearly depends on the gap magnitude. The maximum value of 26 mV 
observed at 2$\Delta$ = 41 meV is in good agreement with in other 
studies.\cite{15} Thus, in cuprates, the phase-coherence energy scale 
$\Delta _c$ has the maximum Josephson strength. As a consequence, the 
dependence $I_cR_n (p)$ has the shape of a dome (see Fig. 12.28 in Ref. 6).

Typical temperature dependences of the two gaps are sketched in Fig. 10a. 
The $\Delta_c$ gap closes with increase of temperature more rapidly than 
that of the BCS-type gap. This can be seen in Fig. 7b. The dependence 
$\Delta_p (T)$ is reminiscent of the BCS one. On heating, $\Delta_p$ closes 
at $T_{pair}$ which is above\cite{20,21} $T_c$. This means that, between 
$T_c$ and $T_{pair}$, $\Delta_p$ makes a contribution to the PG.\cite{21} 
In Bi2212, the pairing $\Delta_p(0)$ gap is proportional to 
$T_{pair}$ as\cite{22,R2} $2\Delta_p(0) \simeq 6\,k_BT_{pair}$. 
\begin{figure}[t]
\centerline{\psfig{file=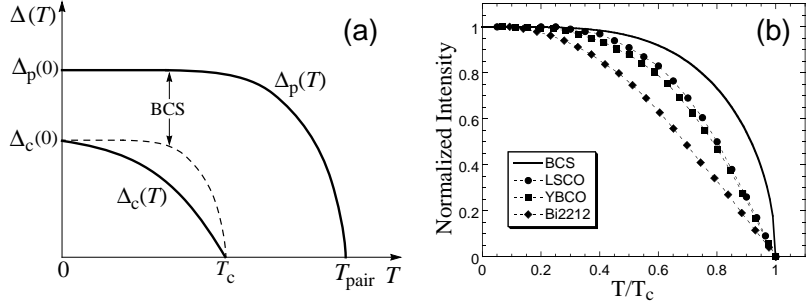,width=10.6cm}}
\vspace*{8pt}
\caption{(a) Typical temperature dependences of $\Delta_c$ and $\Delta_p$ in 
Bi2212.\protect\cite{R,R2} The BCS temperature dependence of the gap 
is shown by the dashed line. The dependence $\Delta_p(T)$ is similar to the 
BCS one. 
(b) Temperature dependences of the peak intensity of the incommensurate 
elastic scattering in LSCO (x = 0) ($T_c$ = 42 K) and the intensity of the 
magnetic resonance peak measured by INS in near optimally doped Bi2212 
($T_c$ = 91 K) and YBCO ($T_c$ = 92.5 K). 
The solid line is the BCS temperature 
dependence.\protect\cite{R,R2,24,25}}
\end{figure}

\section{Magnetic Origin of $\Delta_c$}  

Analysis of several experimental facts indicates that the phase-coherence 
energy gap, $\Delta_c$, has the magnetic origin.\cite{R,R2,17,23,24,25,26,27} 
In other words, the long-range phase coherence in cuprates is mediated by spin 
fluctuations. Consider just a few facts. 
The dependence $\Delta_c(T)$ shown in Fig. 10a is defined by the 
temperature dependence of the intensity of spin fluctuations in cuprates. 
Figure 10b shows temperature dependences of the peak intensity of 
the incommensurate elastic scattering in LSCO and of the intensity of 
the commensurate resonance peak measured by INS in near optimally doped 
Bi2212 and YBCO. In some cuprates, the magnetic excitations in the odd 
(acoustic) channel undergo an abrupt sharpening on cooling through $T_c$. 
This sharp mode in the odd channel is called the magnetic resonance peak, 
which appears at the AF wave vector $Q$ = ($\pi,\pi$) below $T_c$. The 
resonance peak is a manifestation of a collective spin excitation. The
AF correlations weaken in the overdoped region; however, the magnetic 
relaxation is still dominant in the highly overdoped region. In Fig. 10, 
one can see that the $\Delta_c(T)$ dependence in Fig. 10a exhibits 
a striking similarity with those in Fig. 10b. 

At any doping level, the energy $E_r$, at which the resonance peak appears
in INS spectra, equals to 2$\Delta_c$, as shown in Fig. 11a for YBCO, Bi2212 
and Tl2201. To remind, the energy 2$\Delta_c$ is the condensation energy 
of a Cooper pair. 

Recently, $\mu$SR measurements showed that the energy scale $\Delta_c$ 
in cuprates has the magnetic origin. At low doping level and low temperature 
in cuprates, there is a magnetic phase similar to spin glass. The 
dependence $T_g(p)$, where $T_g$ is the transition 
temperature to the spin-glass-like phase, is linear as depicted in 
Fig. 11b for (Ca$_x$La$_{1-x}$)(Ba$_{1.75-x}$La$_{0.25+x}$)Cu$_3$O$_y$ 
(CLBLCO). Decreasing $x$ in CLBLCO, the SC dome shrinks, and the 
magnetic $T_g$ scale falls down too. However, at all $x$, $T_c$ and $T_g$ 
remain ``congruent.'' If one plots $T_c/T^{max}_c$ as a function of 
$K(x)\Delta y$, where $\Delta y = y - 7.15$, and chooses $K(x)$ so that all 
$T_c/T^{max}_c$ domes collapse to a single curve, then, 
$T_g/T^{max}_c$ lines collapse to a single one too.\cite{26} This fact 
indicates that the same single energy scale controls both the SC and 
magnetic transitions. As a consequence, $\Delta_c$ has the magnetic origin. 
The same scaling procedure made for LSCO, YBCO and Bi2212 gives 
identical results.\cite{27} 
\begin{figure}[t]
\centerline{\psfig{file=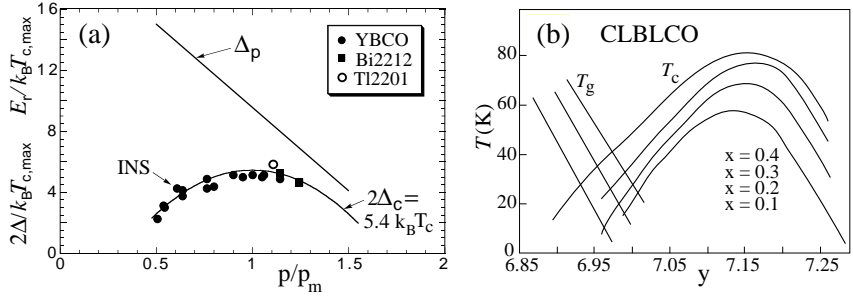,width=11.3cm}}
\vspace*{8pt}
\caption{(a) The phase diagram of cuprates from Fig. 8 and the 
energy position of the magnetic resonance peak, $E_r$, in Bi2212 (squares) 
YBCO (dots) and Tl2201 (circle) at different dopings 
($p_m$ = 0.16).\protect\cite{R,R2,24,25}  
(b) Phase diagram of CLBLCO 
as a function of $y$ at different $x$. The straight lines show 
the spin-glass-like transition temperature, $T_g$, for different 
$x$, determined by $\mu$SR.\protect\cite{26}}
\end{figure}

Since the long-range phase coherence appearing at $T_c$ couples the CuO$_2$ 
layers, $\Delta_c$ is, as a matter of fact, the $c$-axis energy gap. Tunneling 
measurements performed in micron-size mesas, i.e. along the $c$-axis, show 
that the QP peaks have a temperature dependence reminiscent of 
$\Delta_c(T)$ in Fig. 10a and of those in Fig. 10b.\cite{28} Bearing this 
fact in mind, one can draw another important conclusion from Fig. 10, namely, 
that the in-plane mechanism of HTSC (i.e. $\Delta_p$) apparently has no or 
very little relation to magnetic interactions.

\section{``Tunneling'' Pseudogap above and below $T_c$} 

The PG is a depletion of the DOS above $T_c$. Cuprates have a connected 
Fermi surface that appears to be consistent with conventional band theory. 
At all dopings, the PG is pinned to the Fermi level and, therefore, dominates 
the normal-state low-energy excitations. The PG was observed for the first 
time in NMR measurements and, therefore, mistakenly interpreted as a spin 
gap. Later ARPES, tunneling, Raman, specific-heat and infrared measurements 
also provided evidence for a gap-like structure in electronic excitation 
spectra.\cite{4} Thus, it became clear that the PG is not a spin gap but a gap 
to both spin and charge excitations. Alternatively, there are two spatially 
separated PGs: one is a spin gap and the other is a charge gap. There is a 
consensus on doping dependence of PG(s): the magnitude of PG(s) is large in 
the underdoped region and decreases as $p$ increases. 

Consider first tunneling data obtained in Bi2212 above $T_c$. Figure 12a 
shows the temperature dependence of an SIS conductance. In Fig. 12a, there 
is no sign indicating at what temperature the phase-coherence gap was closed. 
Across $T_c$, the spectra evolve continuously into a PG. In Fig. 12a, the 290 K
conductance looks smooth; however, when enlarged vertically, it exhibits 
the presence of a weak PG with the magnitude of about 60 meV 
(see Fig. 12.23 in Ref. 6). The estimated closing temperature of this PG is 
about 350 K. 
\begin{figure}[t]
\centerline{\psfig{file=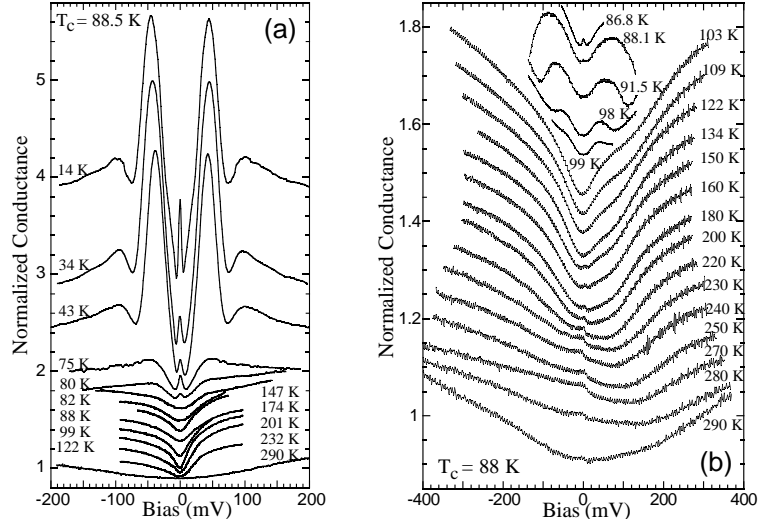,width=10.4cm}}
\vspace*{8pt}
\caption{(a) Temperature dependence of a conductance 
obtained in a Bi2212 single crystal with $T_c$ = 88.5 K ($p \simeq$ 0.19) 
in an SIS junction. The spectra are offset vertically for clarity.
(b) Temperature dependence of a conductance above $T_c$, 
obtained in another Bi2212 with $T_c$ = 88 K by an SIS junction. The 
spectra are offset vertically for clarity.\protect\cite{21,R} 
In both plots, the conductance ``width'' varies because of the 
variations of $R_n$ with temperature.}
\end{figure}

Figure 12b depicts the temperature dependence of another SIS conductance 
exclusively above $T_c$. In Fig. 12b, the spectra are slightly asymmetrical about 
zero bias. The asymmetry is caused, most likely, by the AF medium.\cite{29} 
Analysis of even parts of the conductances, $G_e \equiv [G(V) + G(-V)]/2$, 
shows that the low-bias humps visible in the upper spectra in Fig. 12b 
vanish at about 115 K. The temperature dependence of zero-bias 
conductance has a kink in slope also near 115 K.\cite{R} The low-bias humps 
were interpreted as a contribution from incoherent Cooper pairs.\cite{21} In this 
case, the temperature 115 K is $T_{pair}$ for this doping level of Bi2212. The 
temperature 115 K as well as the interpretation is consistent with other 
tunneling tests.\cite{20} Above 115 K, the spectra in Fig. 12 reflect the 
presence of a normal-state PG. Knowing the values of $T_{pair}$ and 
$\Delta_p$ for this doping level, one can estimate the reduce gap ratio 
($2\Delta_p(0)/k_BT_{pair} \approx 6$). 
Running slightly ahead, it is worth to note that the temperatures 115 K and 
350 K are consistent with one another (see Figs. 13c and 14). 
The temperature dependence of the pairing gap, 
$\Delta_p(T)$, is similar to the BCS one as depicted in Fig. 10a.   

The normal-state PG contributes to tunneling spectra not only above $T_c$ 
but also below $T_c$. The tunneling PG was first seen inside vortex 
cores.\cite{30,R} In the underdoped region, where SC in cuprates is weak, the PG 
was directly observed even in an SIS junction. Figure 13a shows SIS $dI(V)/dV$ 
and $I(V)$ characteristics obtained in an underdoped Bi2212 single crystal with 
$T_c =$ 51 K ($p \simeq$ 0.085). The spectra in Fig. 13a look like usual 
tunneling characteristics for Bi2212. The gap magnitude of 64 
meV is in good agreement with Fig. 8. The $dI(V)/dV$ and $I(V)$ 
characteristics depicted in Fig. 13b are obtained within the {\em same} single 
crystal as those in  Fig. 13a; however, they look differently, and the gap 
magnitude of 130 meV is too large for a SC gap. The conductance humps in 
Fig.13b resemble the humps in Fig. 13a. For some reasons, the QP peaks did 
not develop in the conductance shown in Fig. 13b, uncovering the PG below 
(the sub-gap at low bias is what is left from the QP-peak contribution). 
From these data, one can infer that the humps in conductances 
are caused by a normal-state gap (in fact, the humps in Fig. 13b do not 
correspond {\em directly} to the PG---it is a product of the peak-to-PG 
tunneling, as shown in Fig. 5). Taking into account that these data are obtained 
in SIS junctions, one can conclude that $\Delta _{pg} \simeq 3 \Delta _p$ as 
depicted in Fig. 13c [in SIN junctions, $|V_{hump}| \simeq 3\,|V_{peak}|$ 
(see above)]. Thus, the pairing gap appears on top of (inside) a normal-state 
PG. The result of mathematical subtraction between the two sets of the data 
presented in Figs. 13a and Fig. 13b will therefore correspond to a contribution 
from the SC condensate, and will be discussed below. To conclude, tunneling 
characteristics obtained below $T_c$ in cuprates consist of two 
{\em independent} contributions---from the SC condensate (QP peaks) and 
from the PG (humps at high bias). Tunneling data taken recently in 
micron-size mesas\cite{28} and ARPES data\cite{14} measured in Bi2212 
at momentum near (0, $\pi$) also show that QP peaks and humps in the 
spectra have different origins, and humps are caused by a normal-state PG. 
\begin{figure}[t]
\centerline{\psfig{file=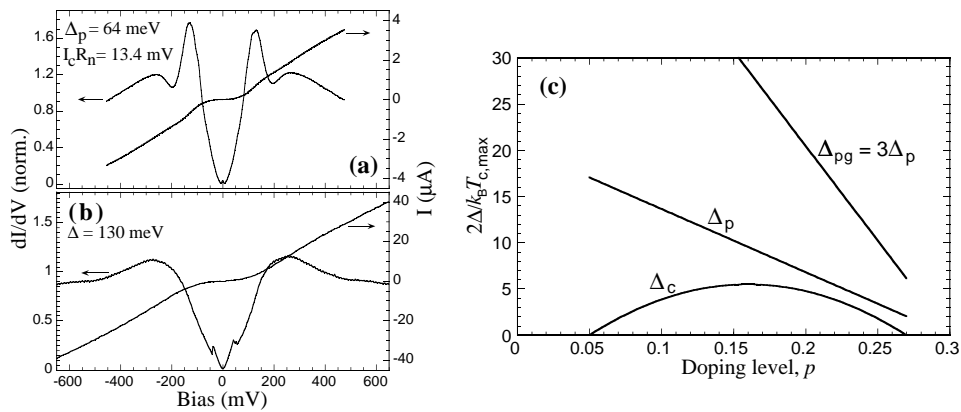,width=12.7cm}}
\vspace*{8pt}
\caption{(a) and (b) SIS tunneling $dI(V)/dV$ and $I(V)$ characteristics 
measured at 14 K within the same underdoped Bi2212 single crystal having 
$T_c$ = 51 K. In both plots, the $dI(V)/dV$ are normalized at -400 mV. 
In plot (a), $I_cR_n$ denotes the Josephson product.\protect\cite{R} 
(c) The phase diagram of SC cuprates from Fig. 8, showing also 
the tunneling normal-state PG, $\Delta_{pg}$.\protect\cite{R,R2}}
\end{figure} 

In the following section, it will be shown that the tunneling PG is most likely 
a charge gap. Analysis of tunneling and ARPES data indicates that the the 
{\em magnitude} of the tunneling PG is not affected much by cooling through 
$T_c$; however, there is a re-organization of excitations {\em inside} the PG 
(see Fig. 12.9 in Ref. 6 or Fig. 5.15 in Ref. 7). Below $T_c$, the SC gaps are 
predominant. Between $T_c$ and $T_{pair}$, incoherent Cooper pairs in 
cuprates contribute to tunneling characteristics.

\subsection{Magnetic Normal-State Pseudogap}

The tunneling PG is not the only one in cuprates. 
In heat-capacity, NMR and transport measurements, the PG vanishes at 
$p \simeq$ 0.19 (a quantum critical point in cuprates), as  depicted in 
Fig. 14 for Bi2212. Elsewhere,\cite{32,R,R2} it was suggested that the 
magnetic $T_{MT}$ and the charge ordering $T_{CO}$ temperature scales 
belong 
to different phases; thus, they are spatially separated. Tunneling surface 
mapping measurements in Bi2212 show that there are patches of, at least, 
two different phases in CuO$_2$ planes.\cite{33} What is interesting 
is that, 
in a first approximation, the two temperature scales, namely, $T_{CO}$ 
and $T_{MT}$ in Fig. 14 intersect the vertical axis in one point, at 
about 980 K. In the phase-separation scenario however, this fact can 
be naturally understood.\cite{32,R,R2} In order to plot the $\Delta_{pg}$ 
energy scale in Fig. 14,  we assumed that $T_{CO} \simeq 3T_{pair}$ 
since $\Delta _{pg} \simeq 3 \Delta _p$. 
The $T_c$ and $T_{MT}$ temperature scales in Fig. 14 have the magnetic 
origin. 
\begin{figure}[t]
\centerline{\psfig{file=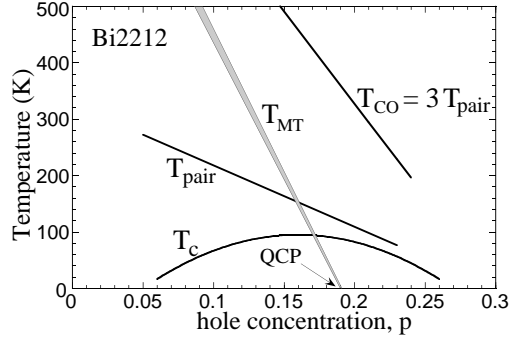,width=6.7cm}}
\vspace*{8pt}
\caption{Phase diagram for Bi2212 based on Fig. 13c. 
$T_{pair}$ is the pairing temperature ($2\Delta_p = 6k_BT_{pair}$), 
$T_{CO}$ is the PG temperature ($T_{CO} = 3 T_{pair}$; CO = charge 
ordering), and $T_{MT}$ is the PG temperature obtained in transport, 
NMR and specific-heat measurements (MT = magnetic 
transition).\protect\cite{32,R,R2,31} QCP is a quantum 
critical point which is located in Bi2212 at $p \simeq$ 0.19.}
\end{figure}

\section{Measurements in Ni and Zn Doped Bi2212} 

By promoting or suppressing magnetic interactions in CuO$_2$ planes, one can 
test the origin of some energy scales in cuprates. One method to affect 
magnetic interactions in CuO$_2$ planes is to substitute a small amount of Cu 
atoms for magnetic or non-magnetic atoms. Ni and Zn are the most suitable 
for this task because they are situated next to Cu in the periodic table 
of chemical elements. Indeed, despite their similar effect on $T_c$ (except 
in YBCO), Ni and Zn doped in CuO$_2$ planes cause very different effects on 
their local environment.\cite{R,R2} 
 
Figure 15 presents three typical conductances obtained in pristine, Ni- 
and Zn-doped Bi2212 single crystals. What are the main changes in 
conductances obtained in Ni- and Zn-doped Bi2212 in respect with those taken 
in pure Bi2212? 
First, in conductances obtained in Zn-doped Bi2212, the ZBCP due to the 
Josephson current is usually smaller than that in Bi2212 conductances. In 
contrast, the ZBCP in conductances obtained in Ni-doped Bi2212 is always 
large. This fact indicates that the phase-coherence mechanism in cuprates 
has the magnetic origin. Second, the distance between humps in conductances 
obtained in Ni-doped Bi2212 is always smaller than that in conductances 
taken in pristine samples. In Zn-doped Bi2212, the distance between humps 
is unchanged. This fact reveals the non-magnetic origin of the ``tunneling'' 
PG, i.e. the ``tunneling'' PG is most likely a charge gap. 
\begin{figure}[t]
\centerline{\psfig{file=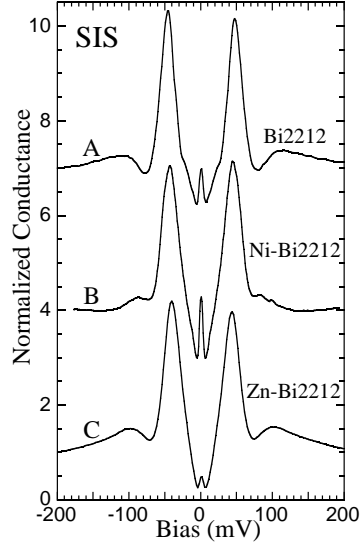,width=4.7cm}}
\vspace*{8pt}
\caption{Normalized conductances obtained by SIS junctions in 
pristine ($T_c \simeq$ 89 K, $p \simeq$ 0.19), Ni-doped ($T_c \simeq$ 75 K) 
and Zn-doped ($T_c \simeq$ 77 K) Bi2212, respectively. The spectra A and B  
are offset vertically for clarity.\protect\cite{34,R}}
\end{figure}

\section{Anomaly in Tunneling $I(V)$ Characteristics}  

We now turn our attention to an anomaly in tunneling $I(V)$ characteristics 
obtained in cuprates. According to the Blonder-Tinkham-Klapwijk (BTK) 
predictions\cite{35} for SIN junctions of conventional SCs, it is anticipated that, 
in the {\em tunneling} regime, $I(V)$ curves at high positive (low negative) 
bias, depending on $R_n$, lie somewhat below (above) normal-state curves. 
In conventional SCs, the BTK predictions are verified by tunneling experiments. 
As an example, Figure 2c shows typical $I(V)$ characteristics of SIN 
junctions for conventional SCs. In cuprates, however, the BTK theory is violated. 
Figure 16a depicts SIS $I(V)$ and $dI(V)/dV$ characteristics measured in 
Bi2212. In Fig. 16a, one can see that the $I(V)$ curve at high 
positive (low negative) bias passes not below (above) the straight line but 
far above (below) the line.\footnote{The dashed line in Fig. 16a is not the 
normal-state curve; however, any straight line passing through zero will show 
that the behavior  of the $I(V)$ curve in Fig. 16a deviates from the 
theory.\cite{35}} 
This fact cannot be explained by the d-wave symmetry of the order parameter. 
We shall refer to this positive offset in $I(V)$ curves as the anomaly. 
\begin{figure}[t]
\centerline{\psfig{file=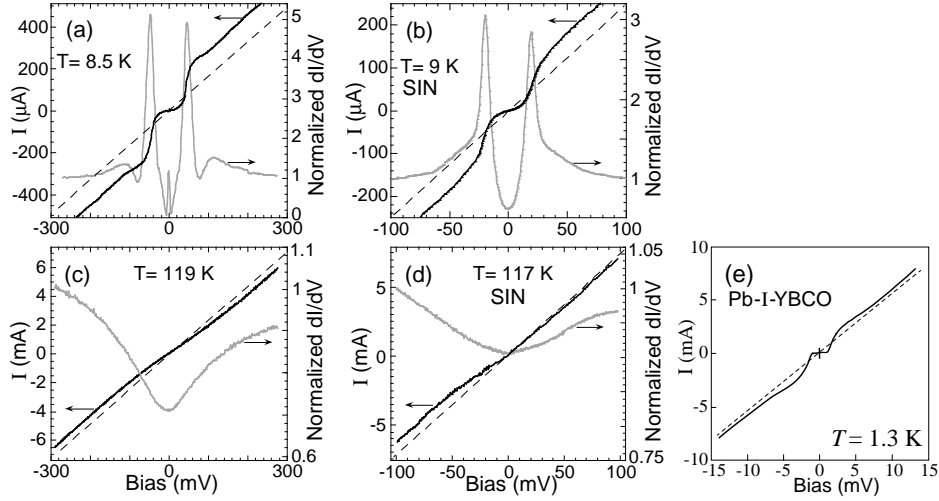,width=12.6cm}}
\vspace*{8pt}
\caption{(a) Tunneling $I(V)$ and $dI(V)/dV$ characteristics obtained 
at $T$ = 8.5 K in an SIS junction of an overdoped Bi2212 single crystal 
with $T_c$ = 88 K. 
(b) $I(V)$ and $dI(V)/dV$ obtained at $T$ = 9 K in an SIN 
junction of an overdoped Bi2212 with $T_c$ = 87.5 K. 
(c) $I(V)$ and $dI(V)/dV$ obtained at 
$T$ = 119 K in the same SIS junction as those in plot (a). 
(d) $I(V)$ and $dI(V)/dV$ obtained at $T$ = 117 K in the 
same SIN junction as those in plot (b).\protect\cite{36,R} 
(e) Tunneling $I(V)$ characteristic taken in an 
in-plane Pb-I-YBCO junction at 1.3 K (I = insulator).\protect\cite{37,R} 
The near optimally doped YBCO has 
$T_c \simeq$ 92 K.  In all plots, the dashed lines are 
parallel to the $I(V)$ curves at high bias.}
\end{figure} 

This anomaly is a manifestation of intrinsic properties of cuprates: it 
appears not only in overdoped Bi2212 but also, as we shall see further, in 
underdoped Bi2212 as well. SIN $I(V)$ curves of Bi2212 exhibit the anomaly too, 
as illustrated in Fig. 16b. The anomaly is also present in $I(V)$ characteristics 
of YBCO, as depicted in Fig. 16e. The anomaly in $I(V)$ characteristics relates 
to $\Delta_p$ and, therefore, vanishes slightly above $T_c$. In Figs. 16c and 
16d, the anomaly is absent, and a small ``negative'' offset from the 
straight line is caused by the PG. 
Thus, the SC condensate gives rise to an anomaly in $I(V)$ characteristics of 
cuprates, on top of a contribution from the PG which has a small ``negative'' 
offset from the straight line. The magnitude of ``negative'' offset scales with 
with the PG magnitude: the larger the PG magnitude is, the larger the 
``negative'' offset is (see Fig. 12.10 in Ref. 6).   

Knowing tunneling characteristics obtained deep below $T_c$ and somewhat 
above $T_c$, and by taking the difference between the spectra, one can 
estimate a contribution in the tunneling spectra from the SC condensate. 
This procedure is equivalent to one used in INS measurements. In our case, 
however, such a subtraction leads only to an estimation of the contribution 
because by subtracting the spectra we assume that the PG crosses $T_c$ 
without modification. However, as discussed above, this is not the case, 
particularly, at low bias. 

In order to compare two sets of tunneling spectra, they should be normalized. 
The conductance curves can be easily normalized at high bias, as illustrated in 
Fig. 17a. How to normalize the corresponding $I(V)$ curves is not a trivial 
question. The conductance curves at high bias, thus, far away from the gap 
structure, are almost constant. Consequently, in a SIS junction, by normalizing 
two conductance curves at high bias, the equation $(dI(V)/dV)_{1,norm} 
\simeq (dI(V)/dV)_{2,norm}$  holds at bias $|V| \gg 2 \Delta_p /e$, where 
$e$ is the  electron charge. By integrating the equation we have 
$I(V)_{1,norm} \simeq I(V)_{2,norm} + C$, where $C$ is the constant, 
meaning that the corresponding $I(V)_{i,norm}$ curves are parallel to each 
other at high bias, as depicted in Fig. 17b.  
\begin{figure}[t]
\centerline{\psfig{file=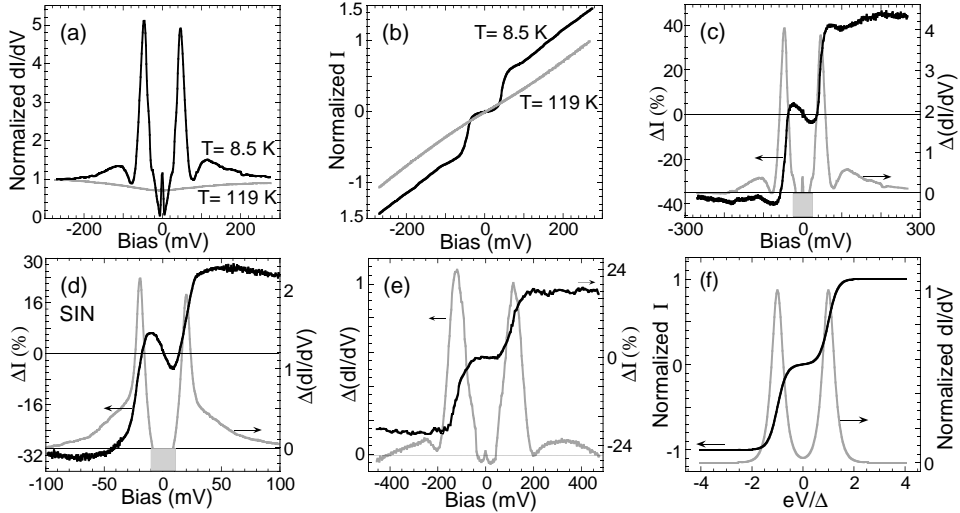,width=12.7cm}}
\vspace*{8pt}
\caption{(a) and (b) Normalized tunneling characteristics from Figs. 16a and 
16c, which are obtained within the same Bi2212 single crystal ($T_c$ = 88 K) at 
different temperatures: (a) conductances, and (b) $I(V)$ curves. 
In plot (b), the normalization procedure is: the normal-state curve is 
normalized by its value at maximum positive bias, and the other curve is 
adjusted to be parallel at high bias to the normalized normal-state curve 
(for more details, see text).   
(c) Differences between two conductances in plot (a) and 
two $I(V)$ curves in plot (b). 
(d) Difference between the two SIN $I(V)$ curves from Figs. 16b and 16d, 
normalized before subtraction as those in plot (b), and the difference 
between their normalized conductances. 
(e) Differences $(dI/dV)_a- (dI/dV)_b$ and $I_a- I_b$ between the two 
sets of spectra shown in Figs. 13a and 13b. The $I(V)$ curves 
were normalized before subtraction as those in plot (b). 
(f) Idealized $I(V)$ characteristic of SC condensate in an SIN junction, and 
its first derivative. The curves are normalized by their maximum values.
In plots (c) and (d), the grey boxes cover the parts of conductances, 
which are below zero and have no physical meaning. In plots (c)--(e), the 
current difference is presented in \%.\protect\cite{R,R2,38,36}}
\end{figure} 

The extracted $I(V)$ and $dI(V)/dV$ characteristics of the SC condensate 
appearing in SIS and SIN junctions of overdoped Bi2212 are presented in 
Figs. 17c and 17d, respectively. Figure 17e shows a {\em low-temperature} 
contribution of the SC condensate to tunneling spectra in {\em underdoped} 
Bi2212. At high bias, the $I(V)$ curves in Figs. 17c--17e reach a plateau value. 
At the gap bias, they rise/fall sharply and, at low bias, the curves go to zero. 
In Figs. 17c and 17d, the negative slope of the $I(V)$ curves at low bias, 
signifying a negative differential resistance, is an artifact which a consequence 
of a rough estimation. Figure 17f depicts idealized $I(V)$ and $dI(V)/dV$ 
characteristics of the SC condensate which summarize the observed 
tendencies. The plateau value in the $I(V)$ curve is small in the underdoped 
region and increases as the doping level rises. 

\section{Electronic States Transfer in Cuprates} 

Let us consider the temperature dependence of the contributions from the 
SC condensate to tunneling spectra of cuprates, shown in Fig. 17f. 
On heating, the plateau value of $I(V)$ curves decreases, and the 
distance between QP peaks in conductances decreases too (see Fig. 6.5 
in Ref. 6 or Fig. 6.45 in Ref. 7). The contributions vanish completely at 
$T_{pair}$. Then, a question naturally arises: where do the states of QP 
peaks go to? Figure 18 shows a sketch of electronic states transfer in 
cuprates at different temperatures, based on tunneling and ARPES data. 
At $T \gg T_c$, there are no in-gap states (except thermal excitations). 
At $T \sim T_c$, there appear mid-gap states transfered from high energies. 
At $T \ll T_c$, the mid-gap states split into two levels, as shown in Fig. 18. 
Interestingly, ARPES spectra which reflect the states exclusively 
above the Fermi level pass the middle step in Fig. 18 unnoticed. Thus, in 
ARPES spectra, the QP-peak states seem to appear directly from high 
energies. 
It is worth to remind that, in contrast to tunneling spectroscopy which 
probe the local DOS, ARPES probes the whole surface of a sample. 
\begin{figure}[t]
\centerline{\psfig{file=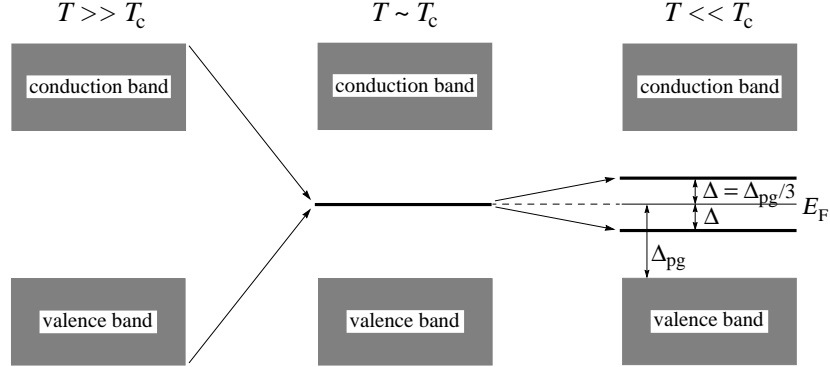,width=10.9cm}} 
\vspace*{8pt}
\caption{Sketch illustrating the electronic states transfer in SC 
cuprates at different temperatures. The gap in the energy spectrum is the 
PG (a charge gap) which is in fact anisotropic. At $T \gg T_c$, there are no 
in-gap states (except thermal excitations). At $T \sim T_c$, there appear 
mid-gap states transfered from high energies 
At $T \ll T_c$, 
the mid-gap states split into two levels: the lower energy level corresponds 
to the bonding one, and the upper one to the antibonding level.
At $T \ll T_c$, the magnitude of pairing gap is about one third of 
PG magnitude, $\Delta = \Delta_p \simeq \Delta _{pg} /3$.\protect\cite{R}}
\end{figure}

\section{Origin of Cooper Pairs in Cuprates} 

The occurrence of in-gap states, as those in Fig. 18, is typical for system 
having topological solitons, for example, in polyacetylene (for a review, see 
Chapter 5 in Ref. 6). Moreover, the $I(V)$ and $dI(V)/dV$ characteristics of 
the SC condensate depicted in Fig. 17f are in good agreement with theoretical 
ones derived for topological solitons.\cite{R,R2,25,38} On the whole, the 
concept is also in agreement with the Davydov bisoliton theory of 
HTSC.\cite{39,40}  
The bisoliton theory utilizes the concept of {\em bisolitons}---electron
(or hole) pairs coupled in a singlet state due to a local deformation of the
lattice. Thus, in the framework of the bisoliton theory, the 
nonlinear electron-phonon interaction is responsible for the QP pairing. 
A possibility for the occurrence of bipolaron SC in quasi-one-dimensional 
(quasi-1D) systems was suggested for the first time by Brazovskii and 
Kirova in 1981.\cite{41} 

Let us consider the derived characteristics for tunneling spectra obtained 
in a system with bisolitons. In such a system, the $I(V)$ characteristic 
obtained in an SIN junction should have the following shape 
\begin{equation} 
c_n(V) = I_0 \times \left[ \tanh \left( \frac{V + V_p}{V_0} \right) + 
\tanh \left( \frac{V - V_p}{V_0} \right) \right], 
\end{equation} 
where $V$ is voltage (bias); $V_p = \Delta_p/e$ is the peak bias, and 
$I_0$ and $V_0$ are constants (``c'' denotes current, and ``n'' an SIN 
junction). Then, the corresponding conductance will have the shape given by 
\begin{equation} 
q_n (V) =  A \times \left[ {\rm sech}^2 \left( 
\frac{V + V_p}{V_0} \right) + {\rm sech}^2 \left( \frac{V - V_p}{V_0} 
\right) \right], 
\end{equation} 
where $A$ is a constant (``q'' denotes QP peaks). In the equations, $V_0$ 
determines the width of conductance peaks. In tunneling spectra, the 
characteristics of bisolitons will appear on top of tunneling characteristics 
of other electronic states present in the system (in our case, a PG).

In SIS junctions, tunneling characteristics represent the convolution of the 
DOS with itself. In this particular case, the convolution cannot be resolved 
analytically. However, as shown elsewhere,\cite{R} in a {\em first} 
approximation, SIS tunneling spectra obtained in a system with bisolitons 
can be fitted by 
\begin{equation} 
c_s(V) = I_0 \times \left[ \tanh \left( \frac{V + 2V_p}{V_0} \right) + 
\tanh \left( \frac{V - 2V_p}{V_0} \right) \right], 
\end{equation}  
\begin{equation} 
q_s (V) =  A \times \left[ {\rm sech}^2 \left( 
\frac{V + 2V_p}{V_0} \right) + {\rm sech}^2 \left( \frac{V - 2V_p}{V_0} 
\right) \right], 
\end{equation} 
where $I_0$, $A$ and $V_0$ are constants (``s'' denotes an SIS junction). 

Let us use these functions to fit tunneling data. Figures 19a--19c depict 
tunneling $I(V)$ characteristics and the corresponding $c(V)$ 
fits. For simplicity, we analyze the $I(V)$ data only at positive bias. 
In Figs. 19b and 19c, the $c_s(V)$ and $c_n(V)$ fits, respectively, are 
adjusted so that the differences between the $I(V)$ curves and the fits 
look similar to the $I(V)$ characteristic of the PG shown in Fig. 19a. 
In Figs. 19b and 19c, the amplitude of the $c(V)$ functions, $I_0$, 
can be changed---this will affect only the scale but not the shape of the 
differences corresponding to the PG. Thus, any $I(V)$ curve 
measured in Bi2212 can be easily resolved into the two components if the 
``SC component'' is interpolated by the corresponding $c(V)$ function.

The conductances and the corresponding $q(V)$ fits are depicted in Figs. 
19d--19f. It is worth to remind that the $q(V)$ functions are assumed to 
fit exclusively the QP peaks, but not the humps.  In Fig. 19, one can find good 
correspondence between the tunneling data and the bisoliton characteristics. 
The numbers used in the $c(V)$ and $q(V)$ functions shown in Fig. 19 can 
be found in Table 12.1 in Ref. 6. 
\begin{figure}[t]
\centerline{\psfig{file=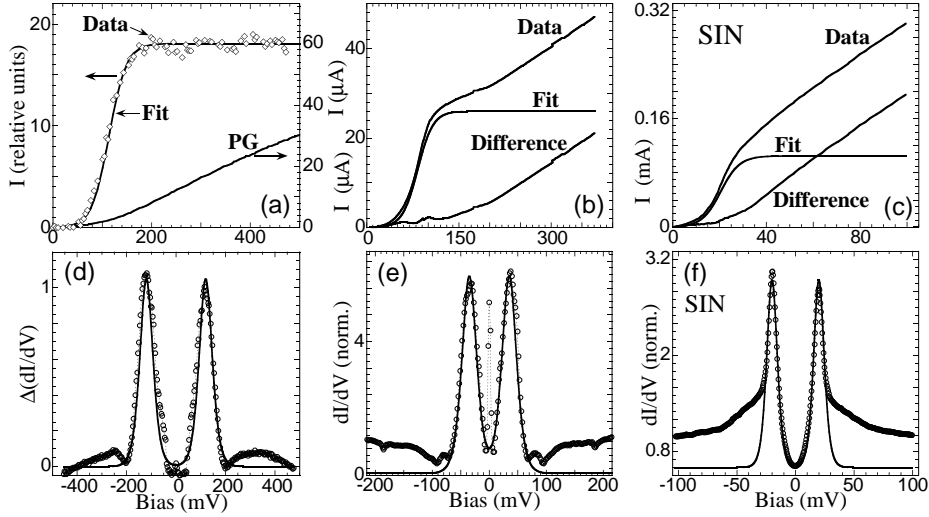,width=12.3cm}}
\vspace*{8pt}
\caption{(a)--(c) $I(V)$ curves obtained in Bi2212 and the $c(V)$ fits (see 
text). (a) SIS $I(V)$ curve (diamonds) from Fig. 17e, the $I(V)$ 
characteristic of PG from Fig. 13b, and the $c_s(V)$ fit. 
(b) SIS $I(V)$ curve obtained in an underdoped Bi2212 single 
crystal with $T_c$ = 83 K (Fig.1 in Ref. 15), the $c_s(V)$ fit, and their 
difference. (c) SIN $I(V)$ curve from Fig. 16b, the $c_n(V)$ fit, and their 
difference. In plots (a)--(c), for simplicity, the $I(V)$ curves are 
presented only at positive bias.  
In plots (b) and (c), the Josephson current at zero bias is not shown. 
(d)--(f) $dI(V)/dV$ curves (circles) obtained in Bi2212 and the 
$q(V)$ fits (see text). (d) SIS conductance from Fig. 17e and the 
$q_s(V)$ fit. 
(e) SIS conductance measured at 15 K in a Ni-doped Bi2212 single crystal 
with $T_c$ = 75 K the $q_s(V)$ fit. 
(f) SIN conductance from Fig. 16b and the $q_n(V)$ fit (see text). 
The spectra in plots (a), (b) and (d) are obtained in underdoped 
Bi2212, while in plots (c), (e) and (f) are measured in overdoped 
Bi2212.\protect\cite{R,R2,25,38}}
\end{figure} 

The presence of soliton-like excitations in cuprates implies the presence 
of quasi-1D in CuO$_2$ planes. There is evidence for charge stripes in 
underdoped LSCO.\cite{42} Transport properties of underdoped YBCO is 
quasi-1D, and there is a kind of charge ordering in undoped 
YBCO, e.g. CDW.\cite{43} On the other hand, the existence of quasi-1D\, QPs 
in cuprates is not a new revelation since induced Cooper pairs on chains 
in YBCO are quasi-1D. This fact is already known  since 1987, after the 
discovery of SC in YBCO. Alternatively, for example, 
discrete breathers which are also topological excitations can exist without 
the presence of quasi-1D in a system (for information about 
discrete breathers, see Section 13 in Chapter 5 of Ref. 6). A possibility for 
the existence of discrete breathers in cuprates was suggested for the first
time in 1996.\cite{44,45} Recently, defuse x-ray scattering measurements 
revealed the existence of lattice modulations favorable for discrete 
breathers.\cite{46,47} The $4a\times4a$ lattice modulation in cuprates 
may occur due to ``frozen'' discrete breathers.

\section{Discussion} 

At the end, let us discuss a few important issues. 
First of all, let us sort out an important question related to the presence of 
two energy gaps in a SC. There exist at least two possibilities for a SC to have 
two energy gaps. First, there are two condensates in a SC, as in MgB$_2$, for 
example. In this case the two gaps are independent, and each of them belongs 
to one of the condensates. Second, there is only one set of Cooper pairs in a 
SC with a
pairing gap $\Delta_p$. They are incoherent above $T_c$. Due to some bosonic 
excitations which are different from those responsible for the pairing, the 
Cooper pairs condense at $T_c$. Thus, for the Cooper pairs, the energy gap 
$\Delta_p$ is ``internal,'' while the phase coherence gap, $\Delta_c$, is 
``external.'' It is easy enough 
to discriminate these two cases. In the first case, the two gaps will appear 
in ARPES spectra, while in the second case, only the pairing gap 
(or $(\Delta _c^2 + \Delta _p^2)^{0.5}$) will be seen by 
APRPES. In cuprates, ARPES detects only $\Delta_p$. This signifies that there 
is only one set of Cooper pairs in cuprates. 

As shown above, spin fluctuations mediate the long-range phase 
coherence in cuprates. This means that $\Delta_c$ has the d$_{x^2 - y^2}$ 
symmetry.\cite{48} According to the third principle of SC presented 
elsewhere,\cite{R,R2} which states that {\em the mechanism of electron 
pairing and the mechanism of Cooper-pair condensation must be different}, 
it is most likely that the electron-phonon interaction is responsible 
for the QP pairing in cuprates. Analysis of some experimental facts points 
out in the same direction.\cite{R,R2} As a reminder, in conventional SCs, 
the two mechanism are different too: 
the electron-phonon interaction leads to the electron pairing, while the 
overlap of their wavefunctions is responsible for the phase coherence. What 
is the symmetry of $\Delta_p$ in this case in cuprates? From ARPES, it is 
strongly anisotropic and, {\em probably}, has an s-wave symmetry. 

What is the main cause of the onset of SC in cuprates? 
After the charge-carrier doping which is absolutely necessary, the 
main cause of the occurrence of SC in cuprates is their 
{\bf unstable lattice}. 
Experimentally, the lattice in cuprates is very 
unstable especially at low temperatures. Upon lowering the temperature, 
all SC cuprates undergo several structural phase 
transitions. In cuprates, the unstable lattice provokes a phase 
separation taking place in CuO$_2$ planes on the nanoscale ($\sim$ a 
few nanometers). Because of a lattice mismatch between different layers 
in {\em doped} cuprates, below a certain temperature in CuO$_2$ 
planes there appear, {\em at least}, two different phases which fluctuate. 
The doped charges prefer to join one of these phases, avoiding the other(s). 
By doing so, the charge presence on one phase enlarges the difference 
between the two phases. Thus, the phase separation is self-sustaining. 
Clusters containing the hole-poor phase in CuO$_2$ planes remain 
antiferromagnetically ordered (at least, up to $p \simeq$ 0.19). This phase 
separation 
taking place in the normal state of cuprates on the nanoscale is the 
main key point for understanding of the mechanism of unconventional SC. 

Spin fluctuations necessary for the phase coherence in cuprates, e.g. 
the so-called magnetic resonance peak, are most likely induced by 
dynamically fluctuating charge stripes and/or discrete breathers. 
Generally speaking, a charge ordering is a manifestation of self-trapped 
states due to the moderately strong electron-phonon interaction. 
Intrinsically, the charge stripes are insulating, i.e. there is a charge gap on 
the stripes. However, if it is the case, the presence 
of soliton-like excitations on the stripes makes them conducting. 

In a {\em first} approximation, unconventional SC in cuprates 
can be described by a combination of two theoretical models: the 
bisoliton\cite{39,40} and spin-fluctuation\cite{49} theories. For example, 
the isotope effect in cuprates is in good agreement with the bisoliton 
theory.\cite{5}

\section{Conclusions} 

To summarize, analysis of the data, 
mainly tunneling, shows that (i) there are two SC gaps in cuprates: a 
pairing and a phase coherence gap. (ii) Tunneling spectra below $T_c$ are a 
combination of coherent QP peaks and incoherent part from a normal-state 
PG. (iii) The ``tunneling'' normal-state PG is most likely a charge gap. (iv) 
However, cuprates have the second normal-state PG which is magnetic 
and, most likely, spatially separated from the charge PG. (v) Just above 
$T_c$, incoherent Cooper pairs contribute to the ``tunneling'' PG.    
(vi) There is a clear correlation between the magnitude of pairing gap and the 
magnitude of ``tunneling'' PG in Bi2212  ($\Delta_p \simeq \Delta_{pg}/3$). 
(vii) The Cooper pairs in cuprates are most likely topological excitations. They 
either reside on charge stripes or represent discrete breathers. 
(viii) The long-range phase coherence occurring at $T_c$ is established due 
to spin fluctuations into CuO$_2$ planes. 
(ix) It is most likely that the nonlinear electron-phonon interaction is 
responsible for the QP pairing in cuprates. 

More details on the mechanism of HTSC based mainly on tunneling 
measurements in cuprates can be found elsewhere.\cite{R,R2,50}

\section*{Acknowledgements} 

I would like to thank W. Y. (Yao) Liang, Peter Littlewood and Gil Lonzarich 
for hospitality at the Cavendish Laboratory.

\end{document}